\documentclass[aps,prl,twocolumn,superscriptaddress,showpacs]{revtex4}
\usepackage{epsfig}
\usepackage{amsmath}
\usepackage{times}

\begin{document}

\title{Detrended fluctuation analysis of traffic data}
\author{Xiaoyan Zhu, Zonghua Liu, and Ming Tang}
\affiliation{Institute of Theoretical Physics and Department of
Physics, East China Normal University, Shanghai, 200062, China}

\date{March 27, 2007}

\begin{abstract}
Different routing strategies may result in different behaviors of
traffic on internet. We analyze the correlation of traffic data
for three typical routing strategies by the detrended fluctuation
analysis (DFA) and find that the degree of correlation of the data
can be divided into three regions, i.e., weak, medium, and strong
correlation. The DFA scalings are constants in both the regions of
weak and strong correlation but monotonously increase in the
region of medium correlation. We suggest that it is better to
consider the traffic on complex network as three phases, i.e., the
free, buffer, and congestion phase, than just as two phases
believed before, i.e., the free and congestion phase.

\end{abstract}
 \maketitle

Tel: 021-62233216(Office); 021-52705019(Home); 13585691004(Cell)\\
Fax:021-62232413\\
Email: zhliu@phy.ecnu.edu.cn\\

Undoubtedly, the internet has become a very important tool in our
daily life. The operations on the internet, such as browsing World
Wide Web (WWW) pages, sending messages by email, transferring
files by ftp, searching for information on a range of topics, and
shopping etc., have benefited us a lot. Therefore, sustaining its
normal and efficient functioning is a basic requirement. However,
the communication in the internet does not always march/go freely.
Similar to the traffic jam on the highway, the intermittent
congestion in the internet has been observed \cite{Huber:1997}.
This phenomenon can be also observed in other communication
networks, such as in the airline transportation network or in the
postal service network. For reducing/controlling the traffic
congestion in complex networks, a number of approaches have been
presented
\cite{arenas1:2002,moreno1:2003,moreno2:2004,moreno3:2004,zhao:2005,
Liu:2006,Liu:2007,Yin:2006,Wang:2006,Echenique:2004,Echenique:2005}.
Their routing strategies can be classified into two classes
according to if the packets are delivered along the shortest path
or not.

The delivering time of a packet from its born to its destination
depends on the status of internet and the routing strategy. It is
believed that there are two phases in communication, i.e., the
free and congestion phase. In the routing strategy of the shortest
path, the delivering time equals the path length in the free phase
and become longer and longer in the congestion phase with time
going. For the former, the delivering times for different packets
will be uncorrelated as each individual packet can go freely to
its destination; while for the later, they will become correlated
as the waiting times are determined by the accumulated packets in
their paths. In the routing strategy of non-shortest path, the
delivering times may be different for different strategies even in
the free phase. As the internet has a power-law degree
distribution, the nodes with heavy links are easy to be the middle
stations for packets to pass by and hence are easy to be
congested. For reducing congestion on these heavy nodes, the
packet may be delivered along a path which avoids the heavy nodes
and hence the path is a little longer than the shortest path
\cite{Echenique:2004,Echenique:2005,Liu:2006,Liu:2007}. Of course,
the packets will still go the shortest path if the packets in the
network is not accumulated. Therefore, it is possible for the
delivering times to be either correlated or uncorrelated in the
free phase. As the delivering times are closely related to the
degree of accumulation of packets in the networks, the correlation
of delivering times can be also reflected in the time series of
packets of the network. A typical routing strategy of the shortest
path is given by Liu {\sl et al.} \cite{Liu:2006}. And two typical
routing strategies of the non-shortest path are given by Echenique
{\sl et al.} \cite{Echenique:2004,Echenique:2005} and Zhang {\sl
et al.} \cite{Liu:2007}. Here we will study the correlation of
packets produced by these three typical strategies.

As the traffic data are produced by all the nodes with some
randomness, there exist erratic fluctuation, heterogeneity, and
nonstationarity in the data. These features make the correlation
difficult to be quantified. A conventional approach to measure the
correlation in this situation is by the detrended fluctuation
analysis (DFA), which can reliably quantify scaling features in
the fluctuations by filtering out polynomial trends. The DFA
method is based on the idea that a correlated time series can be
mapped to a self-similar process by integration
\cite{Peng:1994,Peng:1995,Liu:1999,Yang:2004,Chen:2005}.
Therefore, measuring the self-similar feature can indirectly tell
us information about the correlation properties. The DFA method
has been successfully applied to detect long-range correlations in
highly complex heart beat time series \cite{Peng:1995}, stock
index \cite{Liu:1999}, and other physiological signals
\cite{Chen:2005}. In this paper, we will use the DFA method to
measure the correlation of traffic data.

Most of the previous studies assume that the creation and
delivering rates of packets do not change from node to node.
Considering the fact that different nodes in the internet have
different capacities, a more realistic assumption is that the
packet creation and delivering rates at a node are
degree-dependent. This feature has been recently addressed by Zhao
{\em et al.} \cite{zhao:2005} and Liu {\em et al.}
\cite{Liu:2006,Liu:2007}. They assume that the creation and
delivering rates of packets are $\lambda k_{i}$ and $(1+\beta
k_{i})$, respectively, where $k_i$ is the degree of node $i$,
$\lambda$ represents the ability of creating packets for a node
with degree one, the $1$ in $(1+\beta k_{i})$ reflects the fact
that a node can deliver at least one packet each time, and $\beta$
denotes the ability for a link to deliver packets. For a fixed
$\lambda$, there is a threshold $\beta_c$. It is in the free phase
when $\beta>\beta_c$ and in the congestion phase when
$\beta<\beta_c$. Here we study how the scaling of correlation
changes with the parameter $\beta$. We find that there is always a
scaling in the DFA of traffic data and the scaling can be divided
into three regions, which implies the existence of three phases of
traffic on complex networks, i.e., the free, buffer, and
congestion phase.

\begin{figure}
\epsfig{figure=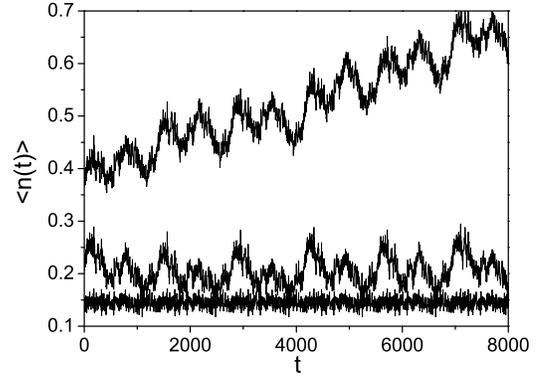,width=0.9\linewidth} \caption{The average
packets per node for $\lambda=0.01$ where the three lines from top
to bottom represent the case of $\beta=0.06, 0.061$, and $0.1$,
respectively.} \label{traffic}
\end{figure}

We now construct a scale-free network with the total number of
nodes $N=1000$ and the average number of links connected with one
node $<k>=6$ according to the algorithm given in Ref.
\cite{LLYD:2002} and let every node create $\lambda k_i$ packets
and send out at most $1+\beta k_i$ packets at each time step. The
destinations of the created packets are randomly chosen and the
sending out obeys the first-in-first-out rule. In the delivering
process, the newly created and arrived packets will be placed at
the end of the queues of each node. For the Liu's approach of the
shortest path, we follow the Ref. \cite{Liu:2006} to collect the
time series of traffic data. Figure \ref{traffic} shows the
evolution of the average packets per node $<n(t)>$ in the network
where the three lines from top to bottom represent the cases of
$\beta=0.06<\beta_c, \beta=0.061\approx \beta_c$, and
$\beta=0.1>\beta_c$, respectively. Obviously, the packets in the
congestion phase of $\beta=0.06$ increase linearly with time $t$,
and the packets in the free phase of $\beta=0.061$ and $0.1$
fluctuate around different constants. For the Echenique's approach
of the non-shortest path, we follow the Ref.
\cite{Echenique:2004,Echenique:2005} that a packet of node $i$
will choose one of its neighboring nodes, $\ell$, as its next
station according to the minimum value of $\delta_{\ell}=h
d_{\ell,j} + (1-h) n_{\ell}$, where $d_{\ell,j}$ is the shortest
path length from the neighboring node $\ell$ to the destination
$j$ and $n_{\ell}$ is the accumulated packets at node $\ell$. The
parameter $h$ is a weighing factor, which can be taken as a
variational parameter and $h \approx 0.8$ is found to give the
best performance. The Echenique's approach thus accounts for the
waiting time only at the neighboring nodes. Echenique's approach
was presented for the case of equal creation and delivering rates
at every node. For the delivery rate of $(1+\beta k_{i})$, a
modified Echenique's approach \cite{Liu:2007} is to choose a
neighboring node with the minimum value of
\begin{equation}\label{eq:Echenique}
\delta_{\ell} = h d_{\ell,j} + (1 - h)\frac{n_{\ell}}{1 + \beta
k_{\ell}}.
\end{equation}
We here choose $h=0.85$ in Eq.(\ref{eq:Echenique}) and find that
the traffic data has the similar behaviors for different $\beta$
as that shown in Fig. \ref{traffic}. And for the Zhang's approach
of the non-shortest path, we follow the Ref. \cite{Liu:2007} to
collect the traffic data. For a packet at node $i$, we take a node
$\ell$ from the neighbors of node $i$ and label the shortest path
from node $\ell$ to the source $j$ by $\{SP:\ell,j\}$. Along this
path, we evaluate the following quantity for the node $\ell$:
\begin{equation}\label{eq:zhang}
d(\ell) = \sum_{s \in \{SP:\ell,j\}} \frac{n_{s}}{1 + \beta k_{s}},
\end{equation}
where the sum is over the nodes along the shortest path
$\{SP:\ell,j\}$, excluding the destination. Thus, $d(\ell)$ is an
estimate of the time that a packet would take to go from node
$\ell$ to the destination $j$ through the shortest path.  The node
$\ell$ with the minimum of $d(\ell)$ will be chosen as the next
station of the packet at node $i$. We find that the traffic data
also has the similar behaviors for different $\beta$ as that shown
in Fig. \ref{traffic}.

\begin{figure}
\epsfig{figure=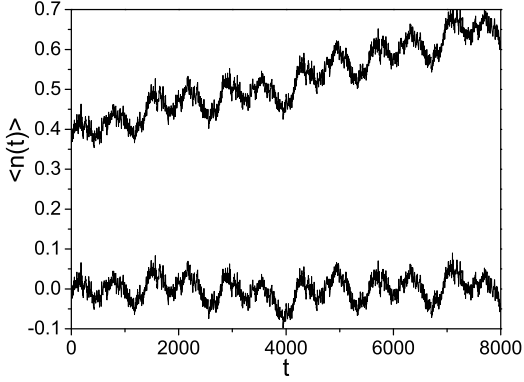,width=0.9\linewidth} \caption{Removing the
global trend of the congestion data for the shortest path approach
where the upper line denotes the original data with
$\beta=0.06>\beta_c$ and the lower line the data after removing
the global trend.} \label{remove_trend}
\end{figure}

All the three typical approaches show a common feature that there
are two kinds of data: the data in the congestion phase increases
linearly with $t$ and the data in the free phase fluctuations
around a constant. In order to quantify the correlations in the
congestion phase, it is important to remove the global trend.
Therefore, we remove the trend of linearly increasing with $t$ by
subtracting a best fitting straight line of the time series. This
procedure makes the data in congestion phase have the similar
behavior with that in the free phase. Figure \ref{remove_trend}
shows an example of removing the global trend where the upper line
denotes the original data with $\beta=0.06>\beta_c$ and the lower
line the data after removing the global trend.

The DFA method is a modified root-mean-square (rms) analysis of a
random walk and its algorithm can be worked out as the following
steps \cite{Peng:1994,Peng:1995,Liu:1999,Yang:2004,Chen:2005}:\\
(1) Start with a signal $s(j)$, where $j=1,\cdots,N$, and $N$ is
the length of the signal, and integrate $s(j)$ to obtain
\begin{equation}\label{eq:step1}
y(i)=\sum_{j=1}^i[s(j)-<s>],
\end{equation}
where $<s>=\frac{1}{N}\sum_{j=1}^Ns(j)$. \\(2) Divide the
integrated profile $y(i)$ into boxes of equal length $m$. In each
box, we fit $y(i)$ to get its local trend $y_{fit}$ by using a
least-square fit. \\(3) The integrated profile $y(i)$ is detrended
by subtracting the local trend $y_{fit}$ in each box:
\begin{equation}\label{eq:step3}
Y_m(i)\equiv y(i)-y_{fit}(i).
\end{equation}
(4) For a given box size $m$, the rms fluctuation for the
integrated and detrended signal is calculated:
\begin{equation}\label{eq:step4}
F(m)=\sqrt{\frac{1}{N}\sum_{i=1}^N[Y_m(i)]^2}.
\end{equation}
(5) Repeat this procedure for different box size $m$. \\
For scale-invariant signals with power-law correlations, there is
a power-law relationship between the rms fluctuation function
$F(m)$ and the box size $m$:
\begin{equation}\label{eq:step5}
F(m)\sim m^{\alpha}.
\end{equation}
The scaling $\alpha$ represents the degree of the correlation in
the signal: the signal is uncorrelated for $\alpha=0.5$ and
correlated for $\alpha>0.5$
\cite{Peng:1994,Peng:1995,Liu:1999,Yang:2004,Chen:2005}.

\begin{figure}
\epsfig{figure=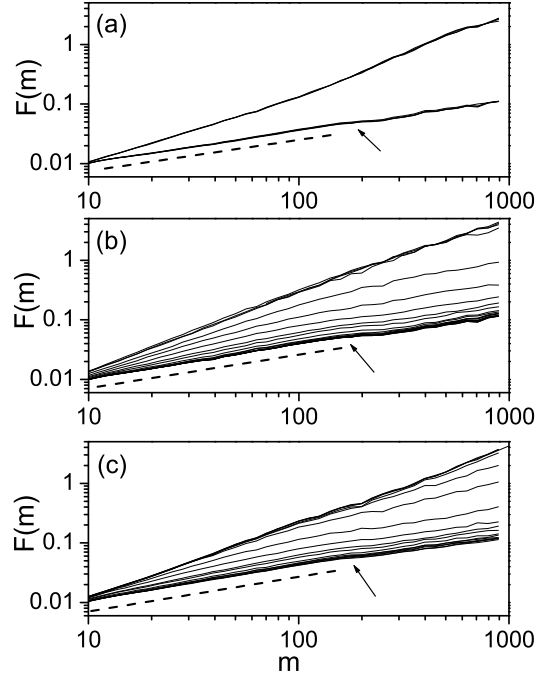,width=0.9\linewidth} \caption{$F(m)$
versus $m$ for different $\beta$ where the arrows show the
crossover points, the dashed lines show the slopes/scalings of
$F(m)$ for eye guide, and (a) represents the case of Liu's
approach, (b) the case of Echenique's approach, and (c) the case
of Zhang's approach.} \label{fig:DFA}
\end{figure}

\begin{figure}
\epsfig{figure=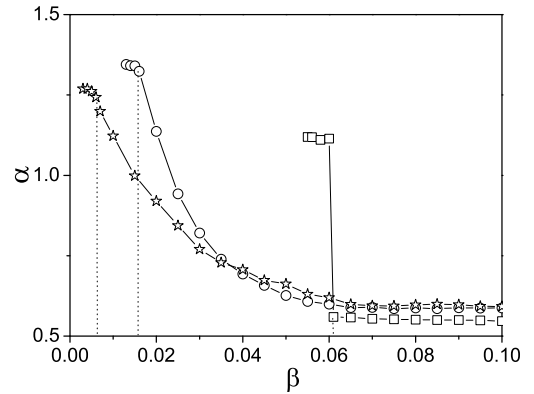,width=0.9\linewidth} \caption{How the
scaling $\alpha$ changes with $\beta$ where the lines with
``squares", ``circles", and ``stars" denote the cases of Liu's,
Echenique's, and Zhang's approach, respectively, and the three
dotted lines show the locations of $\beta_c$ for the three
approaches.} \label{fig:scaling}
\end{figure}

We now use the DFA method to quantify the correlation and scaling
properties of the fluctuated data with no global trend. For the
collected data in the three typical approaches, Fig. \ref{fig:DFA}
shows how the rms fluctuation function $F(m)$ changes with the
scale $t$ where the lines from top to bottom in each panel denote
the direction of increasing $\beta$ and (a) represents the case of
Liu's approach, (b) the case of Echenique's approach, and (c) the
case of Zhang's approach. It is easy to see that all the lines are
straight when $m$ is smaller than the crossover point (shown by
the arrows), indicating there is a scaling $\alpha$ for each line.
Comparing the the lines with different $\beta$, we see that the
scaling $\alpha$ changes with $\beta$. The relationship between
$\alpha$ and $\beta$ is shown in Fig. \ref{fig:scaling} where the
lines with ``squares", ``circles", and ``stars" denote the cases
of Liu's, Echenique's, and Zhang's approach, respectively.

From Fig. \ref{fig:scaling}, it is easy to see that $\alpha$ is an
approximate constant in the congestion phase ($\beta<\beta_c$) for
all the three cases where $\beta_c$ are the locations of the three
dashed lines, but $\alpha$ have different behaviors in the free
phase between the method with the shortest path and that with the
non-shortest path. In the Liu's approach, $\alpha$ in the free
phase is a constant, while in the Echenique's and Zhang's
approaches, $\alpha$ is a constant for $\beta>0.061$ but
monotonously increases before $\beta$ decreases to $\beta_c$.
Let's call the separation value of $\beta$ from a constant to
increasing as $\beta_1$, namely $\beta_1=0.061$. Then, we have
$\beta_1=\beta_c$ in Liu's approach and $\beta_1>\beta_c$ in both
the Echenique's and Zhang's approaches. For $\beta>\beta_1$, all
the values of $\alpha$ are close to $0.5$, hence the corresponding
traffic data are approximately uncorrelated. For
$\beta_c<\beta<\beta_1$, the correlation in the Echenique's and
Zhang's approaches are gradually increased from short range to
long range correlation, and arrive global correlation for
$\beta<\beta_c$. For distinguishing the three different behaviors,
we call their corresponding traffic as {\sl free}
($\beta>\beta_1$), {\sl buffer} ($\beta_c<\beta<\beta_1$), and
{\sl congestion} ($\beta<\beta_c$) phase.

The relationship between the scaling $\alpha$ and the
corresponding traffic behaviors can be understood as follows. In
the strategy of the shortest path, as the accumulation of packets
will firstly occur at the hub nodes, the network can be considered
as congested once the hub nodes are congested \cite{Liu:2006}. In
that time, there are no accumulated packets at the non-hub nodes.
Hence, the critical value $\beta_c$ can be figured out by the
average packets on the hub equalling $1+\beta_c k_{hub}$. Once the
congestion occurs ($\beta<\beta_c$), the accumulation at the hub
nodes will increase linearly and hence influence all the packets
that take the hub nodes as the middle stations. As most of the
shortest paths cross the hub nodes in scale-free networks, the
influence of their congestion will be global and make a
significant change in correlation when the traffic changes from
free to congestion phase. This is why we observe the jump of
$\alpha$ in the line with ``squares" in Fig. \ref{fig:scaling}.

While in the strategy of the non-shortest path, the accumulated
packets at the hub nodes for $\beta_c<\beta<\beta_1$ will not
increase linearly with time even when its average is slightly
larger than $1+\beta k_{hub}$ because the coming packets will
choose other paths to reduce the delivering time. Therefore, for a
fixed creation rate, the packets will go longer and longer paths
to avoid the congestion when the delivering parameter $\beta$
decreases. With the further decrease of $\beta$, more and more
nodes have their average packets larger than $1+\beta k_i$, i.e.,
there are accumulated packets at these nodes. When the nodes with
the smallest links begin to be accumulated with packets, the
congestion occurs. Therefore, the correlation among the packets
will become stronger and stronger with the decrease of delivering
rate $\beta$. That is why we observe the gradually increase of
$\alpha$ in the lines with ``squares" and ``circles" in Fig.
\ref{fig:scaling}. On the other hand, a packet will choose its
path from all the nodes when $\beta<\beta_c$, thus the correlation
will become global. And for $\beta>\beta_1$, the average packets
at the hub nodes will not be over $1+\beta k_{hub}$, but the
fluctuation of packets may be over $1+\beta k_{hub}$ sometimes.
Once it happens, the coming packets will choose other paths to
avoid the accumulation at the hub nodes, resulting a short range
correlation even in the free phase. So we observe the ``circles"
and ``stars" is a little higher than the ``squares" for
$\beta>\beta_1$ in Fig. \ref{fig:scaling}. In sum, a difference
between the buffer phase and the free phase is that the average
packets on a node will be over $1+\beta k_i$ for the former but
not for the later; and a difference between the buffer phase and
the congestion phase is that the average packets on a node will
increase linearly with time in the congestion phase but not in the
buffer phase.

In conclusions, we have investigated the correlation of traffic
data for three typical routing strategies by the DFA method. We
find that there are two phases in the strategy of the shortest
path but three phases in the strategy of the non-shortest path.
The buffer phase comes from the fact that the coming packets will
go a little longer paths with small nodes to avoid the heavy
accumulation at the hub nodes. The average packets in the buffer
phase is larger than that in the free phase but does not increase
linearly with time. The finding of the buffer phase may shed light
on the way of further studying the commulation in internet.

This work was supported by the NNSF of China under Grant No.
10475027 and No. 10635040, by the PPS under Grant No. 05PJ14036,
by SPS under Grant No. 05SG27, and by NCET-05-0424.

\end{document}